\documentclass[3p,times]{elsarticle}
\usepackage{ecrc}

\volume{00}

\firstpage{1}

\journalname{Procedia Computer Science}

\runauth{}

\jid{procs}

\jnltitlelogo{Procedia Computer Science}

\CopyrightLine{2011}{Published by Elsevier Ltd.}

\usepackage{amssymb}
\usepackage[figuresright]{rotating}
\usepackage{multirow} % 引入multirow宏包
\usepackage{graphicx}
\usepackage{subcaption}% 引入多图绘制宏包
\usepackage{siunitx} % 引入输入摄氏度宏包 \SI{25}{\degreeCelsius} 
\begin{document}

\begin{frontmatter}



\dochead{}
%% e.g. \dochead{17th International Conference on Dynamical Processes in Excited States of Solids}

\title{The method to improve the speed of RF switches based on vanadium dioxide}


\author{Tiantian Guo}

\address{Key Laboratory of Hunan Province for 3D Scene Visualization and Intelligence Education(2023TP1038), Hunan First Normal University, No.1015 Yuelu District Fenglin Sanlu,  ChangSha  410205,  China.}

\begin{abstract}
This article proposes a method to improve the switching rate of RF switches based on thermally induced phase change materials. Based on the principle that during the heating process, the increase in heat provided by the heating element plays a major role, while the heat dissipation effect of the bottom heat dissipation layer during the cooling process plays a major role. By replacing the heat dissipation layer material in the phase change RF switch with a high thermal conductivity material, the temperature rise rate of the phase change switch slightly decreases, while the temperature drop rate is significantly increased. Ultimately, the switching speed of the switch instances in this article increased by nearly 28.4\%. The proposal of this method provides a new idea for optimizing the switching rate of RF switches based on thermally induced phase change materials in the future.

\end{abstract}


\end{frontmatter}

% \linenumbers
\newpage
\section{Background}

In existing communication systems, RF switches\cite{2022Conductive}\cite{200816}\cite{1998RF}\cite{ZhangChou-1622}  play a very important role, especially placing higher demands on high-frequency RF switches\cite{	SinghMansour-1741}, which can achieve signal switching\cite{SinghMansour-1685}, modulation\cite{2024A}, and demodulation; At present, common RF switches include solid-state RF switches\cite{2009Maximum}, MEMS mechanical control switches\cite{2001MEMS}, liquid crystal material control switches\cite{2000Alignment}, phase change material control switches\cite{2009Three}, and switches based on graphene materials\cite{2009Microwave} and other new materials. The performance comparison of various types of switches is shown in  Table\ref{tbl:table1 comparison} below. Solid state RF switches have a faster switching speed and mature technology, but their performance is poor in high frequency, especially terahertz band switches; MEMS mechanical control switches have outstanding performance in high-frequency performance, but there are problems such as slow switching speed, high cost, low integration, and short reliability life. Liquid crystal switches are now widely used in the optical frequency band, with low cost. However, they are based on the principle of electric field regulation, which results in poor radiation resistance and slow switching speed of this type of switch; Currently, graphene and phase change materials have gained a lot of research heat in the field of RF switches. These two materials have excellent high-frequency performance parameters as RF switch materials. However, the existing graphene materials are not stable enough in their preparation process, and only a few samples are obtained by hand tearing in the laboratory. Moreover, the maturity of phase change materials is already high during the preparation process, and because their switching principle is based on changes in crystal structure, they have the characteristics of good radiation resistance, long reliability life, and easy integration. Therefore, considering the current situation, phase change material switches have significant advantages, and phase change materials have outstanding performance as candidate materials for RF switches. However, currently, phase change material based switches mainly change their switching state through thermal excitation, and the switching speed is not ideal enough. Therefore, improving the switching speed of phase change material based RF switches has become the most critical issue.\par
Vanadium dioxide material \cite{2012Phase} \cite{2014Advances}\cite{2004Effect}is a phase change material, which has received great attention from researchers in various fields since Morin et al. first reported the phase transition characteristics of VO$_2$ in 1959; Vanadium dioxide has a monoclinic rutile structure\cite{1956Studies} at low temperatures, but as the temperature increases, the crystal structure changes to a tetragonal rutile structure; As the crystal structure changes, the optical and electrical properties of vanadium dioxide undergo abrupt changes\cite{2004Optical}\cite{2004Effects}\cite{1991Optical}. During the phase transition of vanadium dioxide, the process of photoelectric property mutation is reversible and very rapid, and the mutation process can be completed as quickly as ps level time. However, currently the excitation methods for the phase transition characteristics of vanadium dioxide thin films are mostly thermal excitation. Conventional Si and SiO$_2$ materials have low thermal conductivity\cite{2000Molecular}, which slows down the switching speed of vanadium dioxide materials. Therefore, it is necessary to further optimize the structure of vanadium dioxide based RF switches on the current basis to improve the switching speed of such switches.\par
Given that the turn off process takes up a significant amount of time during the two processes of turning on and off phase change material switches. This article optimizes the material deployment in the RF switch structure based on thermally induced phase change materials, replacing low thermal conductivity materials with high thermal conductivity materials, allowing the heat of the heating module to dissipate quickly, accelerating the switch's turn off rate, and increasing the overall switch speed by nearly 28.4\%; Promote the further practical application of VO$_2$ based thermally induced phase change switches.

\begin{table}
	\small
	\caption{Comparison of Various Switches' Performance}
	\label{tbl:table1 comparison}
	\centering
%	\resizebox{\textwidth}{!}{
	\begin{tabular*}{\textwidth}{|c|c|c|c|c|c|c|c|c|}
		\cline{1-9}
		\multicolumn{2}{|c|}{\multirow{2}{*}{Type}}&\multicolumn{3}{c|}{Solid state RF switch}&Mechanical RF switch&\multicolumn{3}{c|}{New material RF switch}\\
			\cline{3-9}
			\multicolumn{2}{|c|}{}&GaAs&PIN&Si&MEMS&Liquid&Graphene&Phase Change material\\
			\cline{1-9}
			\multicolumn{2}{|c}{Speed}&\multicolumn{3}{|c|}{fast}&slow&slow&slow&need to be proved\\
				\cline{1-9}
			\multicolumn{2}{|c}{Cost}&\multicolumn{3}{|c|}{high}&high&low&high&low\\
				\cline{1-9}
			\multicolumn{2}{|c}{Manufacturing process}&\multicolumn{3}{|c|}{Mature}&Mature&Mature&Immature&Mature\\
				\cline{1-9}
			\multicolumn{2}{|c|}{Integration}&High&Low&High&Low&Low&High&High\\
			\cline{1-9}
				\multicolumn{2}{|c}{anti-radiation ability}&\multicolumn{3}{|c|}{Weak}&Weak&Weak&Weak&Strong\\
			\cline{1-9}
		\multirow{3}{*}{RF Performance}&Isolation&\multicolumn{3}{c|}{low}&high&not sure&high&high\\
		\cline{2-9}
	&IL&\multicolumn{3}{c|}{high}&low&not sure&low&low\\		
		\cline{2-9}
	&Power&\multicolumn{3}{c|}{not sure}&high&not sure&not sure&high\\		
		\cline{1-9}			
\multicolumn{2}{|c}{Reliability}&\multicolumn{3}{|c|}{High}&low&high&low&high\\
	\cline{1-9}			
\noalign{\smallskip}%用于在表格的底部和下面的文本之间添加一些垂直空间，这样表格的右侧线和下方的文本看起来不会对齐。这种方法可以在不改变表格结构的情况下实现视觉上的调整。
	\end{tabular*}

\end{table}

\section{Modeling}
\begin{figure}
	\centering
	\includegraphics[height=3cm]{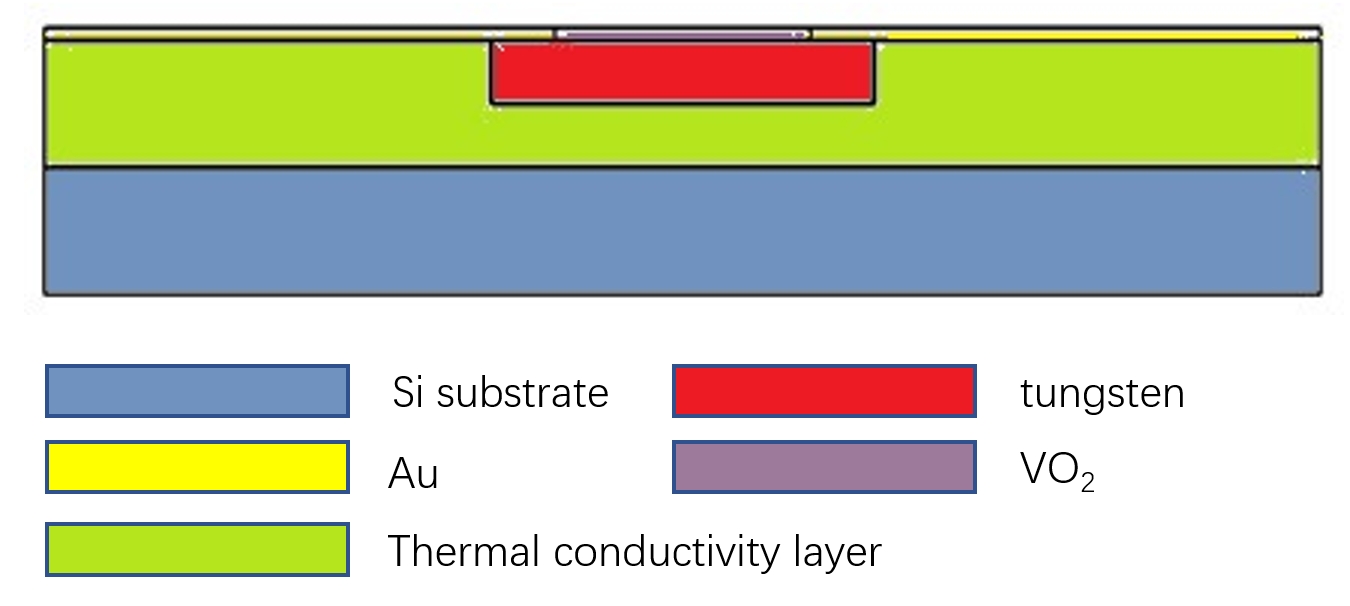}
	\caption{Model of the Switches.}
	\label{fgr:3D Models}
\end{figure}

The structural cross-section of a common vanadium dioxide RF phase change switch based on thermally induced phase transition is shown in Fig\ref{fgr:3D Models}. The top layer consists of a signal line layer and a vanadium dioxide phase change thin film layer, which connects the signal lines made of Au material through the vanadium dioxide thin film. When the vanadium dioxide is in a low temperature state, the signal is disconnected; When vanadium dioxide is in a high temperature state, the signal conducts. The red area at the bottom of the vanadium dioxide film is the tungsten heating layer. By applying voltage to both ends of tungsten, it generates heat and induces a phase transition reaction in the vanadium dioxide film. The green area is the thermal conductive layer substrate layer, and the commonly used substrate layer is SiO$_2$.\par
\begin{figure}
	\centering
	\includegraphics[height=3cm]{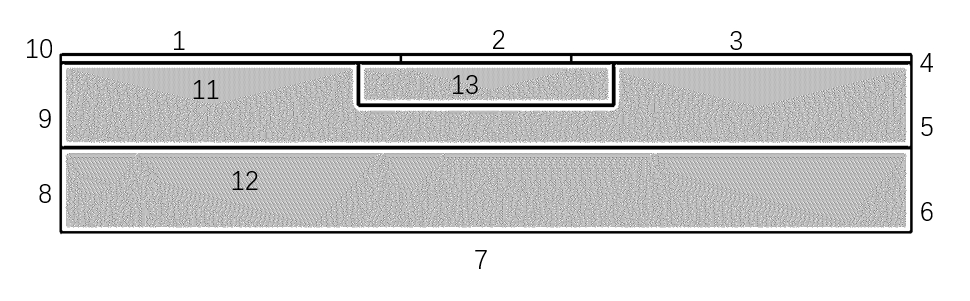}
	\caption{Boundarys of the model.}
	\label{fgr:Boundarys}
\end{figure}

In the simulation model, the initial temperature value of all area materials is set to 291.35 K, and the outer boundaries 1-10 are set as convective heat flux boundaries as shown in Fig\ref{fgr:Boundarys}. The convective heat calculation formula is:

\begin{equation}
	\label{eq:reliang}
{{\rm{q}}_0} = h \cdot ({T_{ext}} - T)
\end{equation}

In the formula, $h$ is the air heat transfer coefficient, set to 5 W/(m$ ^2 $k), ${T_{ext}}$ is the external ambient temperature of 293.15 K, and $T$ is the real-time temperature of the model.\par
Set the tungsten metal in region 13 as the heat source for the model, and define the power of the heat source as the heat consumption rate. Apply a pulse heat source to the model, and the heat source curve is shown in the following Fig\ref{fgr:heatsources}. At 0.1 s, apply a pulse heat source with a pulse power of 0.13 W and a rising transition zone of 0.01  s; Start evacuating the pulse heat source at 0.12 s and evacuate the transition zone for 0.03 s.\par

\begin{figure}
	\centering
	\includegraphics[height=7cm]{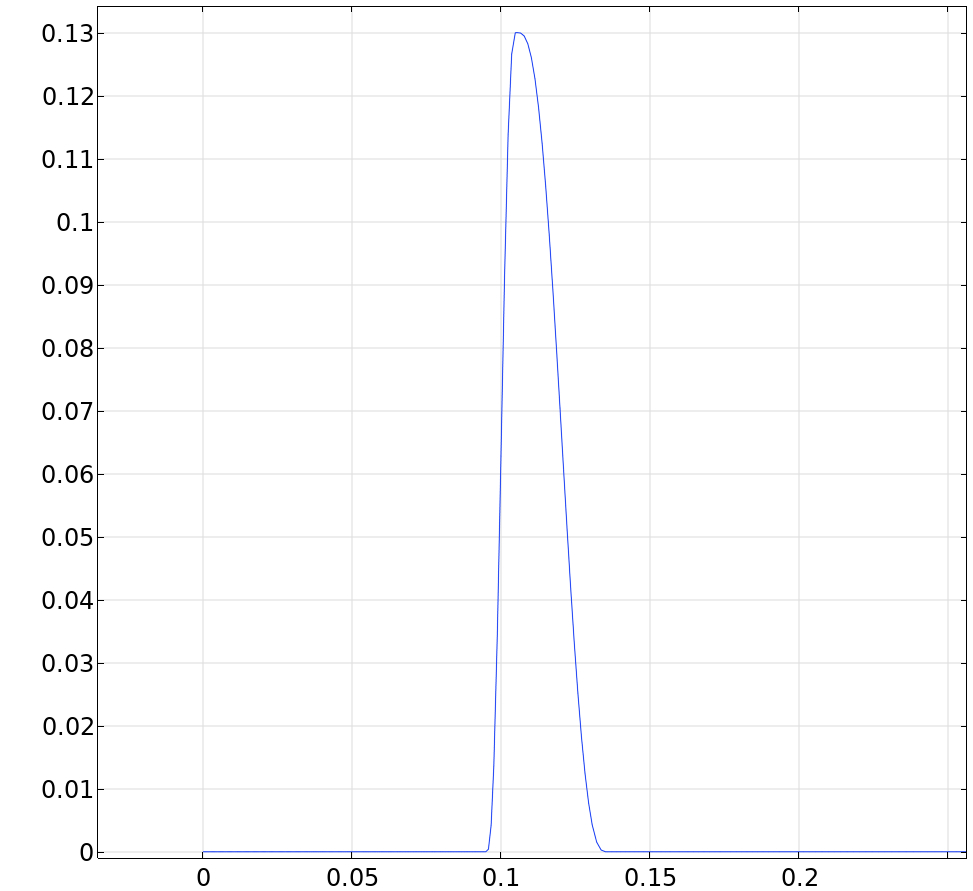}
	\caption{The curve of heat source.}
	\label{fgr:heatsources}
\end{figure}

\begin{figure}
		\centering
		\begin{subfigure}[b]{0.45\textwidth}
			\includegraphics[width=\textwidth]{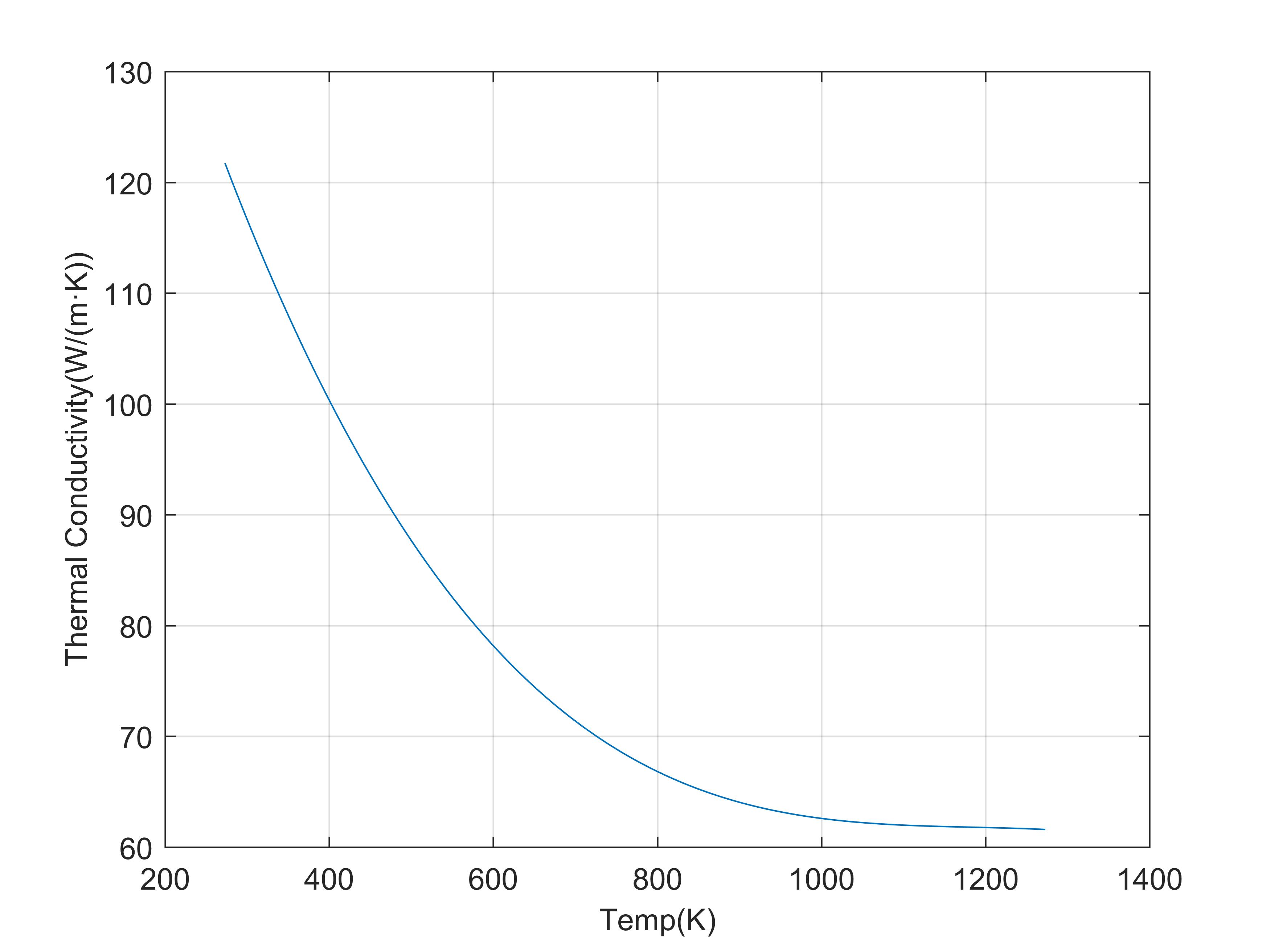}
			\caption{The thermal conductivity of SiC}
			\label{fig:conductivity of SiC}
		\end{subfigure}
		\hfill
		\begin{subfigure}[b]{0.45\textwidth}
			\includegraphics[width=\textwidth]{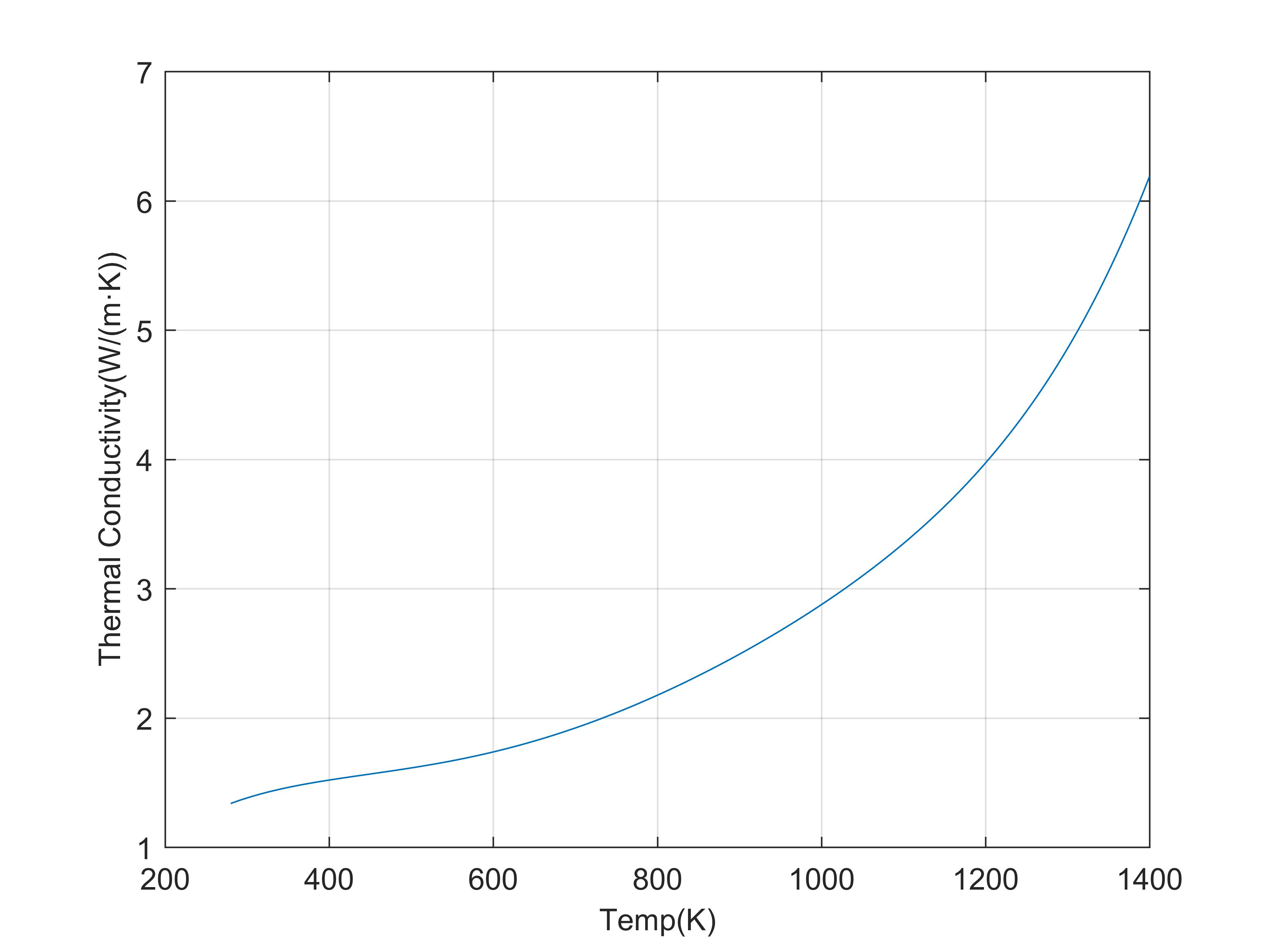}
			\caption{The thermal conductivity of SiO$ _2 $}
			\label{fig:conductivity of SiO2}
		\end{subfigure}
		\caption{thermal conductivity of SiC and SiO$ _2 $.}
	\label{fgr:thermal conductivity }
	\end{figure}

According to the material parameter characteristics, the thermal conductivity of SiO$_2$ is shown on the left side of the Fig\ref{fgr:thermal conductivity }, and the thermal conductivity of SiC \cite{2011Investigation}is shown on the right side of the Fig\ref{fgr:thermal conductivity }. In order to improve the heat dissipation speed of the phase change switch structure, SiC thin film is proposed to replace the thermal conductive layer, which can enhance the heat dissipation speed. As shown in the Fig\ref{fgr:thermal conductivity }, the thermal conductivity of SiC is nearly two orders of magnitude higher than that of SiO$_2$.
According to the difference in thermal conductivity of the materials, two schemes are compared. Scheme one: the material of the thermal conductive layer is set to SiO$_2$; Scheme two: The thermal conductive layer material is set to SiC, a material with high thermal conductivity.

\section{Results}

\begin{figure}
	\centering
	\includegraphics[height=7cm]{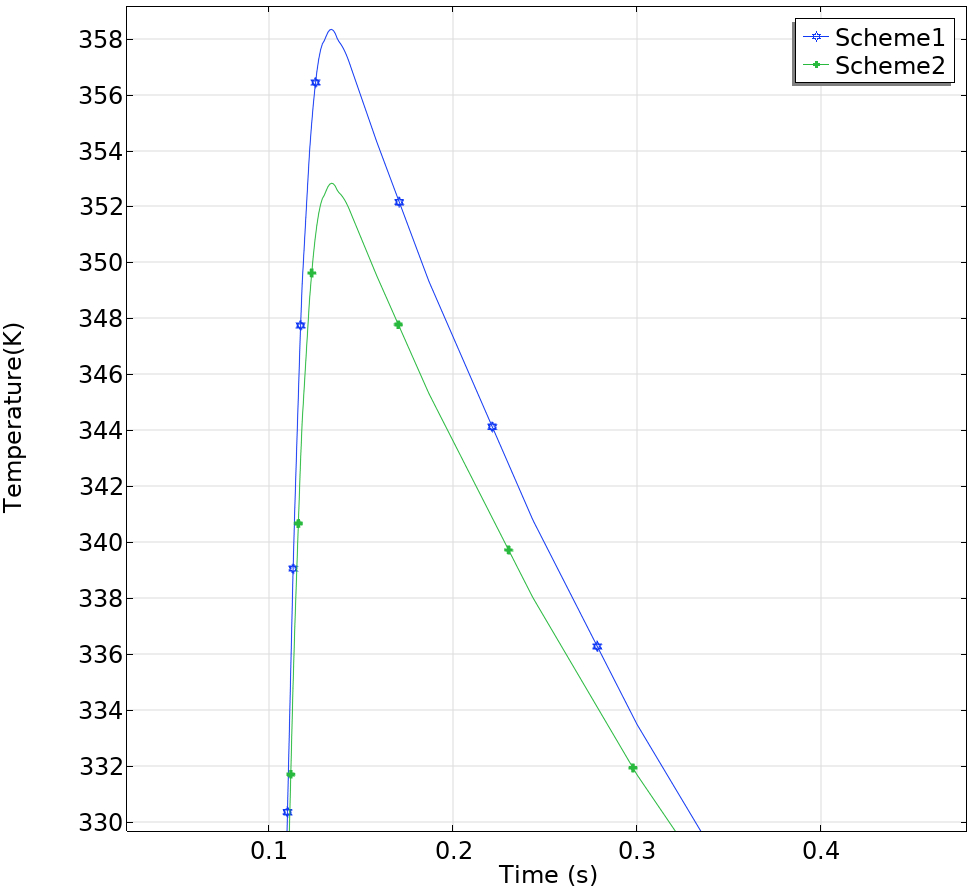}
	\caption{The Tempareture with different schemes.}
	\label{fgr:TempwithTime}
\end{figure}

Solid state heat transfer simulations were conducted on the structures of two schemes, and the temperature at the center of the VO$_2$ film was monitored. The temperature curves of the two schemes over time are shown in Fig\ref{fgr:TempwithTime}. For Scheme 1 using SiO$_2$ for the insulation layer, the temperature at the center of the VO$_2$ film reached 342 K ( \SI{68}{\degreeCelsius}  ) at 0.114 s and continued to rise. When the pulse heat source was removed, the temperature on the film decreased to 342 K at 0.236 s, completing one switch on and off state; For scheme 2 using SiC for the insulation layer, the temperature at the center of the VO$_2$ film reached 342 K ( \SI{68}{\degreeCelsius} ) at 0.117 s and continued to rise. After removing the pulse heat source, the temperature on the film decreased to 342 K at 0.212 s, completing the switching process.\par

\begin{figure}
	\centering
	\begin{subfigure}[b]{0.45\textwidth}
		\includegraphics[width=\textwidth]{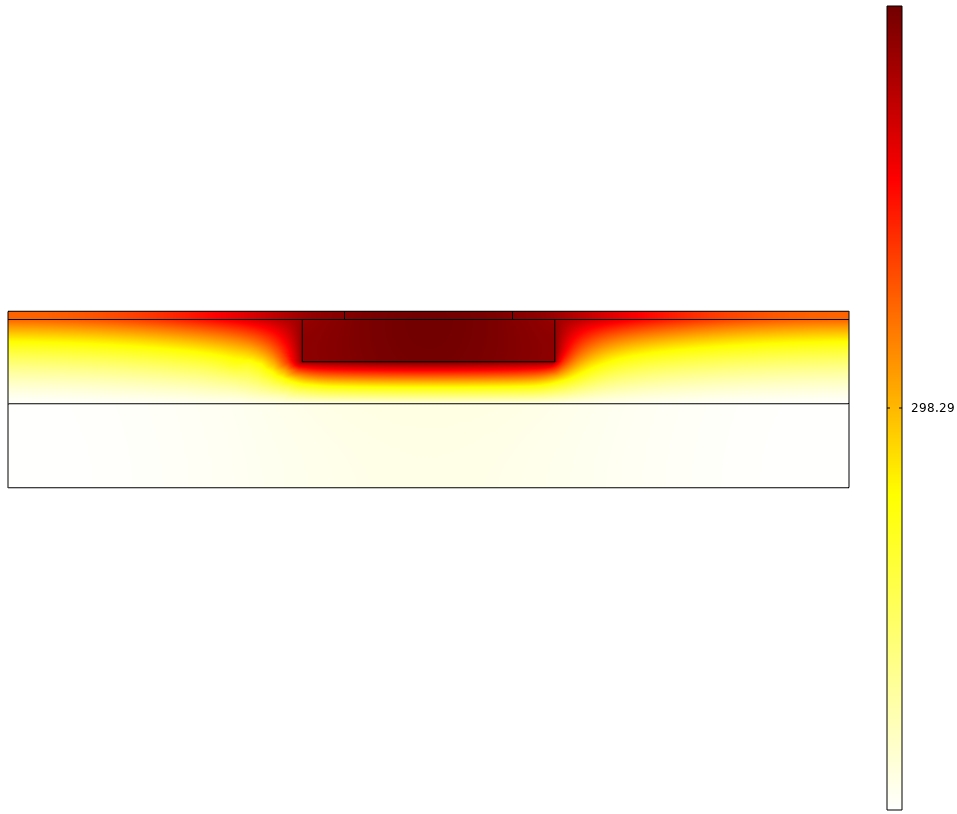}
		\caption{Scheme 1 at 0.1 s}
		\label{fig:TempDistributeat0.1ssolution1}
	\end{subfigure}
	\hfill
	\begin{subfigure}[b]{0.45\textwidth}
		\includegraphics[width=\textwidth]{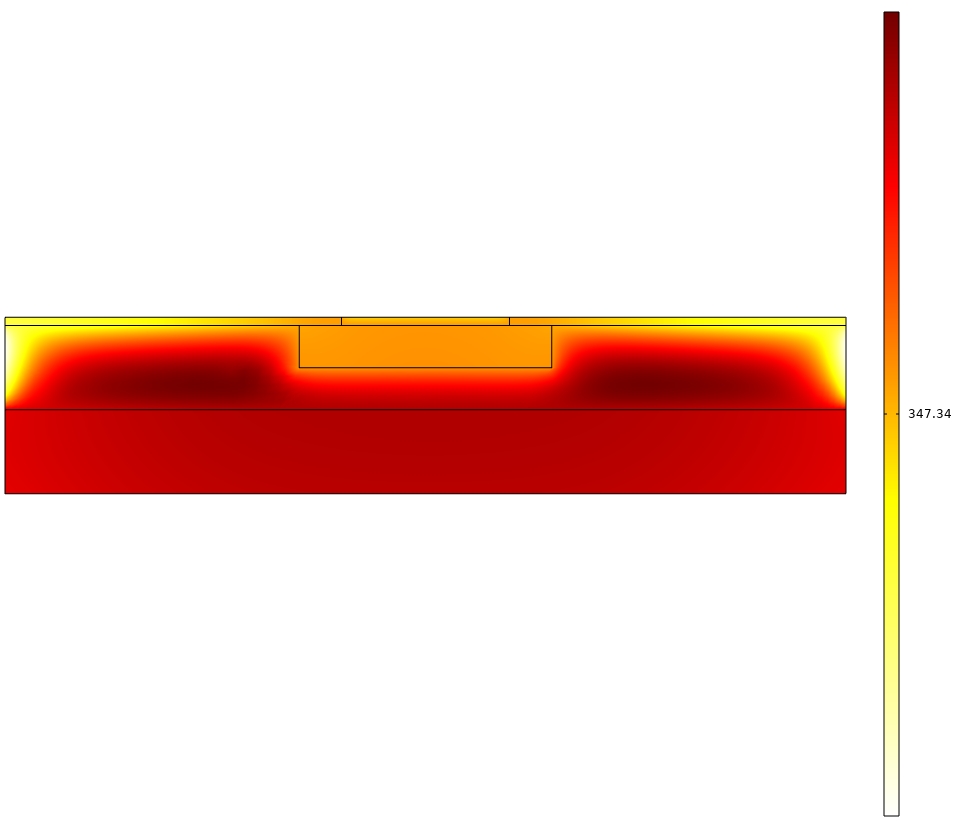}
		\caption{Scheme 1 at 0.2 s}
		\label{fig:TempDistributeat0.2ssolution1}
	\end{subfigure}

\begin{subfigure}[b]{0.45\textwidth}
	\includegraphics[width=\textwidth]{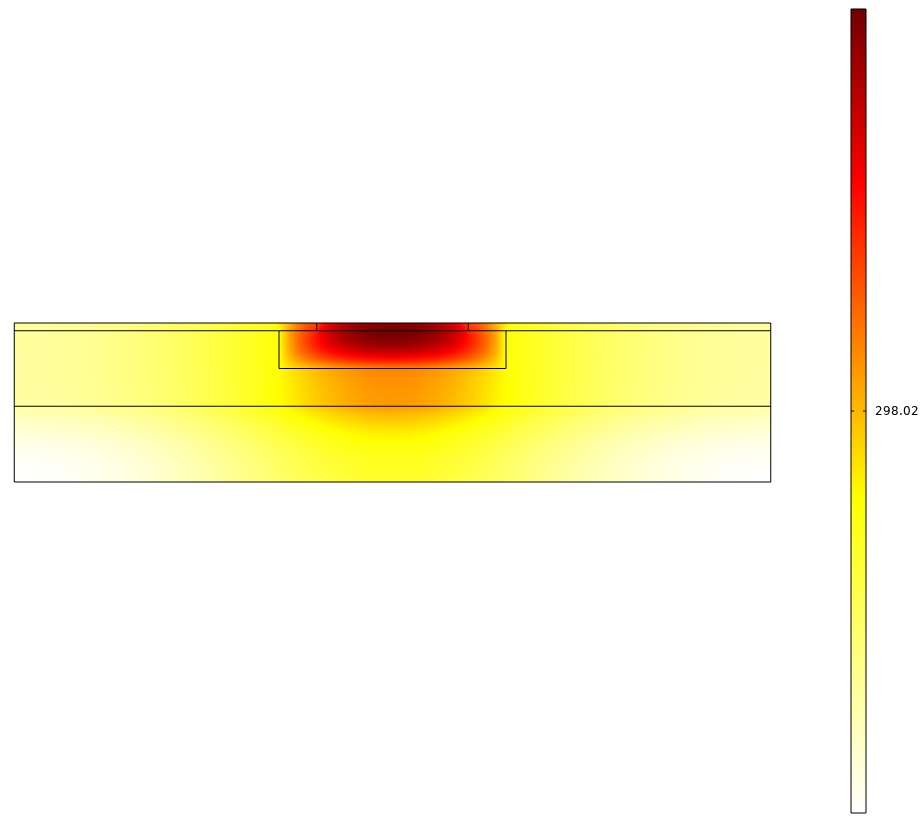}
	\caption{Scheme 2 at 0.1 s}
	\label{fig:TempDistributeat0.1ssolution2}
\end{subfigure}
\hfill
\begin{subfigure}[b]{0.45\textwidth}
	\includegraphics[width=\textwidth]{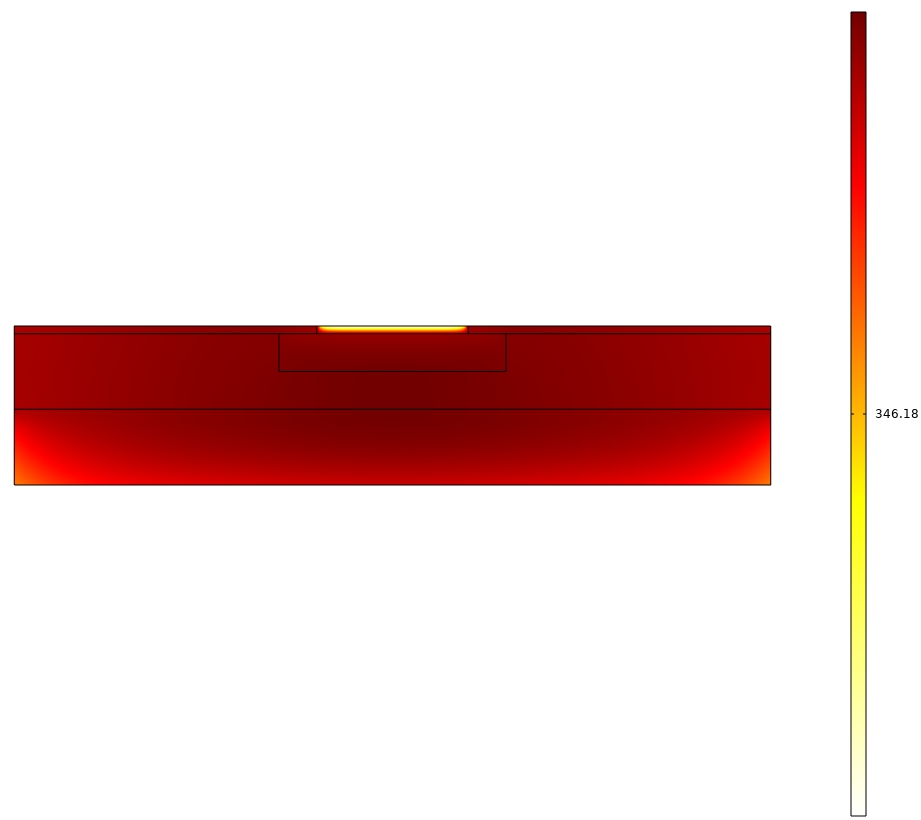}
	\caption{Scheme 2 at 0.2 s}
	\label{fig:TempDistributeat0.2ssolution2}
\end{subfigure}

	\caption{The Temperature Distribution of Model.}
	\label{fgr:Temperature Distribution }
\end{figure}

In the two schemes, the first scheme adopts a low thermal conductivity heat dissipation layer, so in the heating range, heat is preferentially transferred to the vanadium dioxide film above the heating layer, as shown in Fig\ref{fig:TempDistributeat0.1ssolution1}; In the second scheme, the bottom scattering layer is made of a material with high thermal conductivity. Therefore, in the heating range, heat is not only transferred to the upper vanadium dioxide film, but also partially dissipated through the heat dissipation layer, as shown in Fig\ref{fig:TempDistributeat0.1ssolution2}. However, due to the fact that the heat generated by the heating layer during the heating process is much greater than the heat dissipated, the impact of the heat dissipated by the heat dissipation layer on the heating rate of the switch is relatively small. Through simulation calculations, it is found that the switch time is delayed from 0.114 s to 0.117 s, and the switch closing time is delayed by 3 us.\par
In the cooling range, the material with low thermal conductivity in Scheme 2 cannot uniformly diffuse heat, resulting in a time lag in heat dissipation, as shown in Fig\ref{fig:TempDistributeat0.2ssolution1}. In the corresponding scheme one, high thermal conductivity materials are used. From the temperature distribution map of the device, it can be seen that heat is uniformly dissipated through the bottom layer, as shown in Fig\ref{fig:TempDistributeat0.2ssolution2}. By uniformly and efficiently dissipating heat through the bottom heat dissipation layer, the overall cut-off time of the switch is reduced. In Scheme 1, the closing time of the switch is 0.236 s, while in Scheme 2, the closing time of the switch is 0.212 s. Overall, through the optimized solution, the overall switching time of the switch has been increased by 27 us. Of course, the switch simulation examples provided in this article are not the optimal switch design solutions. For RF switches based on thermally induced phase change materials, the switching time of such switches can be further improved by further reducing the switch size or implementing other optimization measures.

\section{Conclusions}

By replacing the heat transfer substrate below with SiO$ _2 $ and high thermal conductivity SiC, the switching speed of the VO$ _2 $  switch based on phase change material was increased from 0.122 s to 0.095 s, with a switching speed increase of 27 us, increase percentage by 28.4\%. The proposal of this method provides a new idea for improving the switching speed of thermally induced phase change RF switches.



%% References


\bibliographystyle{elsarticle-num}
\bibliography{your-bib-database}

\begin{thebibliography}{10}
\expandafter\ifx\csname url\endcsname\relax
  \def\url#1{\texttt{#1}}\fi
\expandafter\ifx\csname urlprefix\endcsname\relax\def\urlprefix{URL }\fi
\expandafter\ifx\csname href\endcsname\relax
  \def\href#1#2{#2} \def\path#1{#1}\fi

\bibitem{2022Conductive}
Z.~R. Xu, Y.~F. Ye, B.~Xia, L.~S. Wu, J.~F. Mao, Y.~Y. Jia, Conductive
  bridging-based memristive rf switches on a silicon substrate, IEEE
  Transactions on Microwave Theory and Techniques~(1) (2022) 70.

\bibitem{200816}
Q.~Li, Y.~P. Zhang, K.~S. Yeo, W.~M. Lim, 16.6- and 28-ghz fully integrated
  cmos rf switches with improved body floating, IEEE Transactions on Microwave
  Theory and Techniques 56~(2) (2008) 339--345.

\bibitem{1998RF}
E.~R. Brown, Rf-mems switches for reconfigurable integrated circuits, IEEE
  Transactions on Microwave Theory and Techniques 46~(11) (1998) 1868--1880.

\bibitem{ZhangChou-1622}
Y.~Zhang, J.~B. Chou, J.~Li, H.~Li, Q.~Du, A.~Yadav, S.~Zhou, M.~Y. Shalaginov,
  Z.~Fang, H.~Zhong, C.~Roberts, P.~Robinson, B.~Bohlin, C.~Ríos, H.~Lin,
  M.~Kang, T.~Gu, J.~Warner, V.~Liberman, K.~Richardson, J.~Hu, Broadband
  transparent optical phase change materials for high-performance nonvolatile
  photonics, Nature Communications 10~(1), item number: 4279 identifier: 12196.
\newblock \href {http://dx.doi.org/10.1038/s41467\-019-12196\-4}
  {\path{doi:10.1038/s41467\-019-12196\-4}}.

\bibitem{SinghMansour-1741}
T.~Singh, R.~R. Mansour, Miniaturized dc–60 ghz rf pcm gete-based
  monolithically integrated redundancy switch matrix using t-type switching
  unit cells, IEEE Transactions on Microwave Theory and Techniques 67~(12)
  (2019) 5181--5190, item number: 8874988.
\newblock \href {http://dx.doi.org/10.1109/TMTT.2019.2944359}
  {\path{doi:10.1109/TMTT.2019.2944359}}.

\bibitem{SinghMansour-1685}
T.~Singh, R.~R. Mansour, Non-volatile multiport dc–30 ghz monolithically
  integrated phase-change transfer switches, IEEE Electron Device Letters
  42~(6) (2021) 867--870, item number: 9416708.
\newblock \href {http://dx.doi.org/10.1109/LED.2021.3076047}
  {\path{doi:10.1109/LED.2021.3076047}}.

\bibitem{2024A}
C.~Wu, C.~Zhu, L.~Zheng, A single sampling double update modulation method to
  enhance low-carrier ratio operation, IEEE Transactions on Power
  Electronics~(2) (2024) 39.

\bibitem{2009Maximum}
Z.~Y. A, X.~H. B, J.~Y. B, G.~S. C, M.~S. A, R.~G. B, Maximum powers of
  low-loss series–shunt fet rf switches, Solid-State Electronics 53~(2)
  (2009) 117--119.

\bibitem{2001MEMS}
T.~W. Yeow, K.~L.~E. Law, A.~Goldenberg, Mems optical switches, Communications
  Magazine IEEE 39~(11) (2001) 158--163.

\bibitem{2000Alignment}
S.~S. Seomun, T.~Fukuda, H.~Matsuda, H.~Miyachi, M.~Kato, Alignment control
  method for liquid crystalline molecules and its application for an
  all-optical device, Applied Physics Letters 77~(1) (2000) 28--30.

\bibitem{2009Three}
H.~Lo, E.~Chua, J.~C. Huang, C.~C. Tan, C.~Y. Wen, R.~Zhao, L.~Shi, C.~T.
  Chong, J.~Paramesh, T.~E. Schlesinger, Three-terminal probe reconfigurable
  phase-change material switches, IEEE Transactions on Electron Devices 57~(1)
  (2009) 312--320.

\bibitem{2009Microwave}
M.~Dragoman, D.~Dragoman, F.~Coccetti, R.~Plana, A.~A. Muller, Microwave
  switches based on graphene, Journal of Applied Physics 105~(5) (2009) 666.

\bibitem{2012Phase}
T.~L. Cocker, L.~V. Titova, S.~Fourmaux, G.~Holloway, H.~C. Bandulet,
  D.~Brassard, J.~C. Kieffer, M.~A.~E. Khakani, F.~A. Hegmann, Phase diagram of
  the ultrafast photoinduced insulator-metal transition in vanadium dioxide,
  Physical Review B 85~(15) (2012) 155120.

\bibitem{2014Advances}
M.~A. Warwick, R.~Binions, Advances in thermochromic vanadium dioxide films,
  Journal of Materials Chemistry A 2~(10) (2014) 3275--3292.

\bibitem{2004Effect}
R.~A. Aliev, V.~A. Klimov, Effect of synthesis conditions on the
  metal-semiconductor phase transition in vanadium dioxide thin films, Physics
  of the Solid State 46~(3) (2004) 532--536.

\bibitem{1956Studies}
G.~Andersson, C.~Parck, U.~Ulfvarson, E.~Stenhagen, B.~Thorell, Studies on
  vanadium oxides. ii. the crystal structure of vanadium dioxide, Acta Chemica
  Scandinavica 10~(12) (1956) 623--628.

\bibitem{2004Optical}
M.~Soltani, M.~Chaker, E.~Haddad, R.~V. Kruzelecky, D.~Nikanpour, Optical
  switching of vanadium dioxide thin films deposited by reactive pulsed laser
  deposition, Journal of Vacuum Science Technology 22~(3) (2004) 859--864.

\bibitem{2004Effects}
M.~Soltani, M.~Chaker, E.~Haddad, R.~Kruzelecky, J.~Margot, Effects of ti–w
  codoping on the optical and electrical switching of vanadium dioxide thin
  films grown by a reactive pulsed laser deposition, Applied Physics Letters
  85~(11) (2004) 1958--1960.

\bibitem{1991Optical}
E.~E. Chain, Optical properties of vanadium dioxide and vanadium pentoxide thin
  films, Applied Optics 30~(19) (1991) 2782--7.

\bibitem{2000Molecular}
S.~G. Volz, G.~Chen, Molecular-dynamics simulation of thermal conductivity of
  silicon crystals, Physical Review B Condensed Matter 61~(4) (2000)
  2651--2656.

\bibitem{2011Investigation}
S.~W. Lee, S.~D. Park, S.~Kang, I.~C. Bang, J.~H. Kim, Investigation of
  viscosity and thermal conductivity of sic nanofluids for heat transfer
  applications, International Journal of Heat and Mass Transfer 54~(1-3) (2011)
  433--438.

\end{thebibliography}






% \end{thebibliography}

\end{document}